\documentstyle[amssymb,preprint,aps,epsf,floats]{revtex}

\tighten
\preprint{\vbox{
\hbox{DOE/ER/40762-262}
\hbox{UMD-PP-03-001}
\hbox{NT@UW-02-022}
}} 
\bigskip \bigskip

\begin{document}
\title{The Chiral Extrapolation of Strange Matrix Elements in the Nucleon}
\author{Jiunn-Wei Chen}
\address{Department of Physics, University of Maryland, \\
College Park, MD 20742\\
{\tt jwchen@physics.umd.edu}}
\author{Martin J. Savage}
\address{Department of Physics, University of Washington, \\
Seattle, WA 98195-1560\\
{\tt savage@phys.washington.edu}}
\maketitle

\begin{abstract}
In current lattice simulations of nucleon properties, 
the up and down quark masses
are significantly larger than their physical values, while 
the strange quark can be included in simulations with its
physical mass.
When the up and down quark masses are much smaller
than the strange-quark mass the chiral extrapolation of 
strange-quark matrix elements in the nucleon from
the lattice up and down 
quark masses to their physical values can be performed
with two-flavor chiral perturbation theory, 
thereby avoiding the slow convergence 
problem of the three-flavor chiral expansion. 
We explore the chiral expansion of several matrix elements of
strange operators in the nucleon
in two-flavor chiral perturbation theory and two-flavor 
partial-quenched chiral perturbation theory.
\end{abstract}

\vfill\eject

The strange quark, $s$, contribution to the properties of the nucleon 
provides a direct probe of the non-valence structure of the nucleon.
If $s$-quarks were significantly more massive than $\Lambda_{\rm QCD}$,
as are the $c$- and $b$-quarks, 
their contributions could be determined within the framework of 
perturbative QCD, 
and could be simply incorporated in the coefficients of operators
involving up, $u$, and down, $d$, quarks and gluons, $g$.
However,  the $s$-quarks are light enough so that their contribution to 
nucleon properties are important and they should be considered as dynamical
degrees of freedom in the nucleon,
but heavy enough so that precise experimental measurements and 
the development of a  rigorous theoretical description
are challenging.

Currently, several $s$-quark properties of the nucleon have been determined
experimentally, including the contribution to the nucleon mass, $m_{s}\langle
N|\overline{s}s|N\rangle $ \cite{sigterm}, the $s$-quark helicity
contribution to the proton spin, $\Delta s$~\cite{ds}, the $s$-quark
contribution to the proton magnetic moment, $\mu _{s}$~\cite{sample}, 
and the unpolarized $s$-quark parton distribution~\cite{sdis}. 
Further, ongoing experiments are expected to map out the
strange electric and magnetic form factors~\cite{sample,HAPPEX,PVA4,G0}
in the near future.
For 
the determinations of $m_{s}\langle N|\overline{s}s|N\rangle $ and 
$\Delta s$, an expansion about the SU(3) flavor symmetry limit, in which 
the $u$-, $d$- and $s$-quarks are degenerate, is used 
but,  unfortunately, in the absence of  further experimental inputs,
higher order SU(3) symmetry breaking effects~\cite{su3b} 
can only be estimated 
with hadronic models or with 
the leading non-analytic contributions from kaon loops 
computed in three-flavor chiral perturbation theory ($\chi$PT).
To accommodate the sometimes slower than expected convergence of
the SU(3) expansion in the baryon sector, 
conservative error estimates inevitably lead to
large uncertainties.
Lattice simulations of the $s$-quark 
properties of the nucleon, and the presently needed extrapolation to 
the physical quark masses must also deal with this issue.
$\chi $PT~\cite{ChPT,GL,HBChPT,ChPTUlf} 
and its quenched- and partially-quenched extensions,
quenched $\chi $PT (Q$\chi $PT)~\cite{QChPT,Qmartin} and 
partially-quenched $\chi $PT (PQ$\chi $PT)~\cite{PQChPT,PQmartin,BS}, 
respectively,
with both two and three flavors have been widely used to extrapolate 
physical quantities simulated on the lattice. 
One has confidence that for small enough quark masses
both the two-flavor and three-flavor theories converge.
However, in the three-flavor theory 
the relatively large value of the $s$-quark mass
means that the issue of convergence must be explored
on  an observable-by-observable basis.
If the SU(3) expansion is found  not to converge rapidly 
it cannot provide a reliable extrapolation of the lattice calculations. 
However, the chiral extrapolations
of strange matrix elements in the nucleon computed in 
lattice simulations do not suffer from the slow convergence of 
SU(3) $\chi $PT, and the 
two-flavor chiral expansion is sufficient for chiral
extrapolations of these observables.
The
reason is very simple---one does not need to know the $s$-quark mass
dependence for the purpose of chiral extrapolation. 
In lattice simulations
only the $u$- and $d$-quark masses are heavier than their physical values,
while
the $s$-quark can be simulated with its physical mass. 
Any variations in the $s$-quark mass can be implemented as an interpolation,
rather than an extrapolation.

In SU(2) $\chi $PT, the small expansion parameter of the theory is 
\begin{equation}
\epsilon \ \sim\ {m_\pi\over\Lambda}\ ,\ {\Delta\over\Lambda}\ ,\ 
{p\over\Lambda}
\ \ \ ,
\label{eps}
\end{equation}
where in nature the physical pion mass is $m_{\pi }\sim 140$ MeV, the
$\Delta$-nucleon mass difference is $\Delta \sim 300$ MeV, 
and $p$ is
the characteristic momentum transfer in any given process.
$\Lambda $
is the ``high-energy'' scale set by the energy gap to strange hadronic
excitations.  A naive estimate of 
$\Lambda \sim m_{K}+m_{\Sigma }-m_{N}\sim 760$ MeV,
(accidentally close to the $\rho $ meson mass)
roughly reflects the energy cost to create a $s\overline{s}$ pair. 
Thus $\epsilon \sim 0.4$ in the real world SU(2) expansion for low energy
processes, consistent with the discussions found in 
Ref.~\cite{ChPTUlf}. 
In the chiral extrapolation of strange matrix elements, the $\epsilon$ in
eq.~(\ref{eps}) also sets the radius of
convergence of two-flavor PQ$\chi $PT  to be $\sim 760$~MeV. 
One expects the theory to converge rapidly for 
``$m_{\pi }$''$\lesssim 300$~MeV.

For
exploratory studies, several $s$-quark properties have been 
computed in
quenched simulations~\cite{QQCDs}. 
As pointed out in Ref. \cite{jw}, the
fact that large-$N_{c}$  quenched QCD (QQCD)
($N_c$ being the number of colors)
and large-$N_{c}$ QCD are identical implies that differences in the low-energy
constants of these two theories are ${\cal O}(1/N_{c})$. 
Thus after including quenched chiral logarithms, which are formally higher
order in $1/N_{c}$ but numerically enhanced by the
smallness of light quark masses, one can hope that a QQCD calculation
will have a ${\cal O}(1/N_{c})\sim 30\%$ error. 
However, for cases such as the
proton strange magnetic moment, the higher order $1/N_{c}$ corrections 
are likely to
be important due to accidental cancellations at
leading order (LO). 
In this work we present the chiral expansion
of several nucleon strange matrix elements up to the
next-to-next-to-leading order (NNLO) (i.e. ${\cal O}(\epsilon ^{2})$) 
in both $\chi $PT and PQ$\chi $PT \cite{PQChPT,PQmartin,BS}. 
PQ$\chi $PT is presently the only tool that can rigorously remove
(partial) quenching errors without performing fully unquenched simulations. 
We perform the PQ$\chi $PT calculations using two
non-degenerate light-quark masses in the hope of providing an extra ``handle''
to aid in the extraction of strange matrix elements in the nucleon 
from lattice data.

\section{Strange Matrix Elements in QCD}

\subsection{Spin Singlet Operators}

We start by considering the matrix element of the scalar current 
$\overline{s}s$ in $\chi $PT. 
This quark level operator is matched to the most 
general sum of hadronic operators 
in the heavy baryon chiral lagrangian~\cite{HBChPT}
which transform as spin and isospin singlets 
\begin{equation}
\overline{s}s\rightarrow A_{S}\ \overline{N}N\ +B_{S}\overline{T}^{\alpha
}T_{\alpha }
\ +\ \cdots 
\ \ \ ,  
\label{ss}
\end{equation}
where $N$ and $T_{\alpha }$ are the nucleon and $\Delta$-resonance fields,
respectively,
and where only the LO operators are shown. 
The prefactors can be
rewritten as matrix elements of the quark operator~\cite{CJ1},
and 
by taking the nucleon matrix elements at LO  we
obtain, $A_{S}=$ $\langle N|\overline{s}s|N\rangle ^{0}$,
and $B_{S}=-\langle \Delta |\overline{s}s|\Delta \rangle ^{0}$,
where the superscript denotes LO. 
At next-to-leading-order (NLO),there is a contribution proportional to 
$\Delta/\Lambda $. 
\begin{figure}[!ht]
\centerline{{\epsfxsize=4.0in \epsfbox{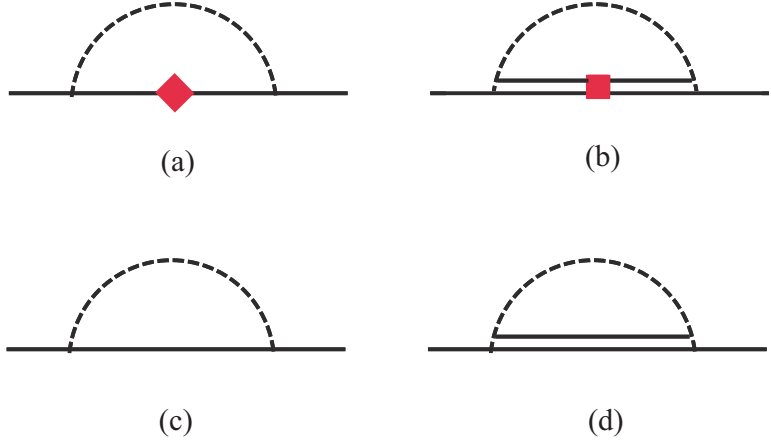}}} 
\vskip 0.15in
\noindent
\caption{\it 
One-loop graphs that give contributions of the form 
$m_q \log m_q$
to strange matrix elements in the nucleon.
A single solid line denotes a nucleon, a double solid line denotes a 
$\Delta$-resonance and a dashed line denotes a pion.
Diagrams (a) and (b) are vertex corrections while diagrams (c) and (d) 
denote wavefunction renormalization.
}
\label{fig:loops}
\vskip .2in
\end{figure}
The NNLO contribution comes from the one-loop diagrams shown in 
Fig.~\ref{fig:loops}
which depend on both $A_{S}$ and $B_{S}$ operators, and tree-level
counterterms. Thus up to NNLO (${\cal O}(\epsilon ^{2})$), 
and working in the isospin limit,
\begin{eqnarray}
\langle N|\overline{s}s|N\rangle & =& 
\langle N|\overline{s}s|N\rangle ^{0}-\ {
\frac{g_{\Delta N}^{2}}{4\pi ^{2}f^{2}}}\left( \langle N|\overline{s}
s|N\rangle ^{0}-\langle \Delta |\overline{s}s|\Delta \rangle ^{0}\right)
J_{\pi }
\nonumber\\
& & \ +\ 
C_{S}\frac{\Delta }{\Lambda }+D_{S}\frac{\Delta ^{2}}{\Lambda
^{2}}+E_{S}\frac{\overline{m}}{\Lambda}
\ \ \ .  
\label{ss2}
\end{eqnarray}
where $g_{\Delta N}=1.8$ is the $\Delta N \pi$ coupling, 
$f\sim 132~{\rm MeV}$ is the pion
decay constant, $\overline{m}=m_u=m_d$ is the 
quark mass in the isospin limit.
The function 
$J_{\pi }$=$J(m_{\pi},\Delta ,\mu )$ is
\begin{eqnarray}
J(m,\Delta ,\mu ) &=&\left( m^{2}-2\Delta ^{2}\right) \log \left( {\frac{
m^{2}}{\mu ^{2}}}\right) +2\Delta \sqrt{\Delta ^{2}-m^{2}}\log \left( {\frac{
\Delta -\sqrt{\Delta ^{2}-m^{2}+i\epsilon }}{\Delta +\sqrt{\Delta
^{2}-m^{2}+i\epsilon }}}\right)  \nonumber \\
&&\ +\ {\frac{5}{3}}(m^{2}-2\Delta ^{2})
\ \ \ ,
\end{eqnarray}
where $\mu$ is the renormalization scale.
The scale dependence of the counterterms $C_S$, $D_S$ and $E_S$ exactly
compensate the scale dependence introduced by the loop function, $J_\pi$.

An analogous result is found for the matrix elements of the
$s$-quark spin-independent twist-2 operators 
\begin{equation}
{\cal O}_{s}^{\mu _{1}\cdots \mu _{n}}=\overline{s}\gamma ^{(\mu
_{1}}iD^{\mu _{2}}\cdots iD^{\mu _{n})}s\ ,  \label{Os}
\end{equation}
where $D^\mu$ is the QCD covariant derivative and where 
($\cdots $) indicates the 
symmetrization and removal of traces of the enclosed
indices. 
We will first focus on
forward matrix elements and leave the off-forward case to the next
section. The operator ${\cal O}_{s}^{\mu _{1}\cdots \mu _{n}}$
is matched to hadronic operators 
\begin{eqnarray}
{\cal O}_{s}^{\mu _{1}\cdots \mu _{n}} & \rightarrow & 
\alpha _{s}^{(n)}\overline{
N}v^{(\mu _{1}}\cdots v^{\mu _{n})}N
\ +\ \beta _{s}^{(n)}\overline{T}^{\alpha
}v^{(\mu _{1}}\cdots v^{\mu _{n})}T_{\alpha }
\nonumber\\
& & +\ \gamma _{s}^{(n)}\overline{T}
^{(\mu _{1}}v^{\mu _{2}}\cdots v^{\mu _{n-1}}T^{\mu _{n})}
\ +\ \cdots ,
\end{eqnarray}
where $v^\mu$ is the nucleon four-velocity.
$\ \alpha _{s}^{(n)}$ and the combination 
$\beta_{s}^{(n)}-\gamma _{s}^{(n)}/3$ can be rewritten as the unpolarized
nucleon and $\Delta$ twist-2 matrix elements, i.e.
$\alpha _{s}^{(n)}=\langle N|{\cal O}
_{s}^{\mu _{1}\cdots \mu _{n}}|N\rangle ^{0}$ and $\beta _{s}^{(n)}-\gamma
_{s}^{(n)}/3=-\langle \Delta |{\cal O}_{s}^{\mu _{1}\cdots \mu _{n}}|\Delta
\rangle ^{0}$. 
The nucleon matrix element of ${\cal O}_{s}^{\mu _{1}\cdots \mu _{n}}$ is
related to the n-th moment of the $s$-quark parton distribution 
\begin{equation}
\left\langle N\left| {\cal O}_{s}^{\mu _{1}\cdots \mu _{n}}\right|
N\right\rangle =2\left\langle x^{n-1}\right\rangle _{s,N}P^{(\mu _{1}}\cdots
P^{\mu _{n})}\ ,\ (n=1,2,3,...)
\ \ \ ,
\end{equation}
where $P^\mu$ is the nucleon momentum, and where
\begin{equation}
\left\langle x^{n-1}\right\rangle _{s}=\int_{0}^{1}dx\ x^{n-1}\left( s\left(
x\right) +\left( -1\right) ^{n}\overline{s}\left( x\right) \right) \ ,
\end{equation}
with $s$ ($\overline{s}$) is the quark (antiquark) spin-averaged
distribution. The variable (Feynman) $x$ is the momentum fraction of the
proton carried by a quark in the infinite momentum frame, and for simplicity
we have suppressed the renormalization scale dependence. To NNLO, 
\begin{eqnarray}
\left\langle x^{n-1}\right\rangle _{s,N}
& = & \left\langle x^{n-1}\right\rangle
_{s,N}^{0}-\ {\frac{g_{\Delta N}^{2}}{4\pi ^{2}f^{2}}}\left( \left\langle
x^{n-1}\right\rangle _{s,N}^{0}-\left\langle x^{n-1}\right\rangle _{s,\Delta
}^{0}\right) J_{\pi }
\nonumber\\
& & \ +\ 
C_{x_S}^{(n)}\ \frac{\Delta }{\Lambda }
\ +\ D_{x_S}^{(n)}\ \frac{\Delta ^{2}}{\Lambda^{2}}
\ +\ E_{x_S}^{(n)}\ \frac{\overline{m}}{\Lambda}
\ \ \ ,
\end{eqnarray}
for $n>1$.  The $n=1$ moment is not renormalized and remains zero, 
as it corresponds to the strangeness charge operator.
It is no surprise that the chiral structure of $\left\langle
x^{n-1}\right\rangle _{s,N}$ is identical to the chiral 
structure of $\left\langle x^{n-1}\right\rangle _{u+d,N}$ 
that was found in Refs.~\cite{AS,CJ} and to 
that of $\langle N|\overline{s}s|N\rangle $ in eq.(\ref{ss2})
as these are all isosinglet and spin-singlet operators.

\subsection{Spin Non-Singlet Operators}

The discussions of the previous section also apply to s-quark 
operators that are not spin singlets.
We start with the proton strange magnetic moment induced by the strange
vector current $\overline{s}\gamma ^{\mu }s$. This operator can be matched
to tree level hadronic operators
\begin{equation}
\overline{s}\gamma ^{\mu }s
\rightarrow 
\frac{1}{m_{N}}\varepsilon ^{\mu \nu
\alpha \beta } v_{\alpha }\partial _{\nu }\left( \mu _{s,N}^{0}\ 
\overline{N}S_{\beta }N\ -3\mu _{s,\Delta }^{0}\ \overline{T}^{\sigma }S_{\beta
}T_{\sigma }\right) 
+\ \cdots ,
\end{equation}
where $S_{\beta }$ is the covariant spin operator, and where 
$\mu _{s,N}^{0}$ and $\mu _{s,\Delta}^{0}$ are the LO 
strange magnetic moments of the nucleon and $\Delta $, respectively.
At NNLO, (${\cal O}(\epsilon ^{2})$), a straightforward calculation reveals
\begin{eqnarray}
\mu _{s,N} & = & \mu _{s,N}^{0}\ -\ {\frac{1}{8\pi ^{2}f^{2}}}\left[ \ \mu
_{s,N}^{0}\left( 3g_{A}^{2}L_{\pi }\ +\ 2g_{\Delta N}^{2}J_{\pi }\right) 
\ -{\frac{10}{3}}\mu _{s,\Delta }^{0}g_{\Delta N}^{2}J_{\pi }\ \right] 
\nonumber\\
& & \ +\ 
C^{(\mu_S)}\frac{\Delta }{\Lambda }+D^{(\mu_S)}\frac{\Delta ^{2}}{\Lambda
^{2}}+E^{(\mu_S)}\frac{\overline{m}}{\Lambda}
\ \ ,
\end{eqnarray}
where $g_A\sim 1.26$ is the $\pi N$ coupling constant, 
and $L_{\pi }=m_{\pi}^{2}\log \left( {m_{\pi }^{2}/\mu ^{2}}\right)$.

Consider now the matrix elements
\begin{eqnarray}
\left\langle N\left| \overline{s}\gamma ^{(\mu _{1}}\gamma _{5}iD^{\mu
_{2}}\cdots iD^{\mu _{n})}s\right| N\right\rangle &=&
2\left\langle
x^{n-1}\right\rangle _{\Delta s,N} M_N
r^{(\mu _{1}}P^{\mu _{2}}\cdots P^{\mu
_{n})}\ ,  \nonumber \\
\left\langle N\left| \overline{s}\sigma ^{\alpha (\mu _{1}}\gamma
_{5}iD^{\mu _{2}}\cdots iD^{\mu _{n})}s\right| N\right\rangle
&=&
2\left\langle x^{n-1}\right\rangle _{\delta s,N} M_N r^{[\alpha }P^{(\mu
_{1}]}P^{\mu _{2}}\cdots P^{\mu _{n})}\ ,  \nonumber \\
~~~~(n &=&1,2,3,...)
\end{eqnarray}
where $\left[ \cdots \right] $ means that the indices enclosed are
antisymmetric and traceless. 
$r^{\mu }$ is the nucleon polarization vector with $r^{2}=-1$. 
These matrix elements are related to
moments of parton distributions 
\begin{eqnarray}
\left\langle x^{n-1}\right\rangle _{\Delta s} &=&\int_{0}^{1}dx\ x^{n-1}\left(
\Delta s\left( x\right) +\left( -1\right) ^{n-1}\Delta \overline{s}\left(
x\right) \right) \ ,  \nonumber \\
\left\langle x^{n-1}\right\rangle _{\delta s} &=&\int_{0}^{1}dx\ x^{n-1}\left(
\delta s\left( x\right) +\left( -1\right) ^{n}\delta \overline{s}\left(
x\right) \right)
\end{eqnarray}
where $\Delta s$ $(\Delta \overline{s})$ is the helicity distribution
, and 
$\delta s$ $(\delta \overline{s})$ is the transversity 
distribution~\cite{filippone}
of the s-quarks (s-antiquarks). 
$\left\langle x^{n-1}\right\rangle _{\delta s,N}$ 
and 
$\left\langle x^{n-1}\right\rangle _{\Delta s,N}$ 
have the same chiral structure as $\mu_s$, and up to NNLO 
\begin{eqnarray}
\left\langle x^{n-1}\right\rangle _{\Delta s(\delta s),N} &=&\left\langle
x^{n-1}\right\rangle _{\Delta s(\delta s),N}^{0}
\nonumber\\
& & 
\ -\ {\frac{1}{8\pi ^{2}f^{2}
}}\left[ \left\langle x^{n-1}\right\rangle _{\Delta s(\delta s),N}^{0}\left(
3g_{A}^{2}L_{\pi }\ +\ 2g_{\Delta N}^{2}J_{\pi }\right) 
\ -\ {\frac{10}{3}}\left\langle x^{n-1}\right\rangle _{\Delta s(\delta
s),\Delta }^{0}g_{\Delta N}^{2}J_{\pi }\ \right] 
\nonumber\\
& & \ +\ 
C_{\Delta s(\delta s),N}\ \frac{\Delta }{\Lambda }
\ +\ D_{\Delta s(\delta s),N}\ \frac{\Delta ^{2}}{\Lambda
^{2}}
\ +\ E_{\Delta s(\delta s),N}\ \frac{\overline{m}}{\Lambda}
\ \ .
\label{ds}
\end{eqnarray}
However, it is important to note that not
all spin non-singlet twist-2 matrix elements have this chiral structure.

The $s$-quark contribution to the proton spin $J_{s}$ is related to the
non-forward matrix element of the strange quark (traceless) energy momentum
tensor ${\cal T}_{s}^{(\mu _{1},\mu _{2})}$ \cite{Ji} which is just the
operator shown in eq.(\ref{Os}) with $n=2$, 
\begin{equation}
{\cal T}_{s}^{(\mu _{1},\mu _{2})}=\overline{s}\gamma ^{(\mu _{1}}iD^{\mu
_{2})}s\ .
\end{equation}
This operator matches to~\cite{Jq} 
\begin{eqnarray}
{\cal T}_{s}^{(\mu _{1},\mu _{2})} 
&\rightarrow &
2v^{(\mu _{1}}\varepsilon^{\mu _{2})\nu \alpha \beta }
v_{\alpha }\partial _{\nu }\left(
J_{s,N}^{0}\ \overline{N}S_{\beta }N\ -3J_{s,\Delta }^{0}\ \overline{T}
^{\sigma }S_{\beta }T_{\sigma }\right)
\ +\ \left\langle x\right\rangle _{s,\pi }^{0}\partial ^{(\mu _{1}}\pi
^{a}\partial ^{\mu _{2})}\pi ^{a}
\ \cdots\  ,
\end{eqnarray}
in the hadronic theory, 
where $J_{s,N(\Delta )}^{0}$ is the LO $s$-quark contribution
to the nucleon ($\Delta$) spin, $\left\langle x\right\rangle _{s,\pi }^{0}$ is
the LO momentum fraction of the pion carried by strange quarks
and $\pi ^{a}$ is the pion field ($a=1,2,3$). 
It is straightforward to show that~\cite{Jq} at NNLO, 
\begin{eqnarray}
J_{s,N} &=&J_{s,N}^{0}\ +\ 
C_{J_S}\frac{\Delta }{\Lambda }+D_{J_S}\frac{\Delta ^{2}}{\Lambda
^{2}}+E_{J_S}\frac{\overline{m}}{\Lambda}
\nonumber\\
& & \ -\ {\frac{1}{8\pi ^{2}f^{2}}}\left[ (J_{s,N}^{0}-
\frac{\left\langle x\right\rangle _{s,\pi }^{0}}{2})\left( 3g_{A}^{2}L_{\pi
}\ +\ 2g_{\Delta N}^{2}J_{\pi }\right)
-{\frac{10}{3}}(J_{s,\Delta }^{0}-\frac{\left\langle x\right\rangle
_{s,\pi }^{0}}{2})g_{\Delta N}^{2}J_{\pi }\ \right] 
\ \ \ .
\label{eq:qcdspin}
\end{eqnarray}

\section{Strange Matrix Elements in PQQCD}

In this section, we compute some of the quantities discussed in the previous
section  in PQ$\chi $PT. The method for computation in PQ$\chi$PT
is well documented in Refs.~\cite{PQChPT,PQmartin,BS} especially in  the
case of two light, non-degenerate flavors~\cite{BS}.
In PQQCD, the fermion sector is generalized from two light quarks to four
fermionic quarks (two ``valence'' quarks and two ``sea'' quarks) and two
bosonic quarks (``ghosts''). 
The lagrangian has a $SU(4|2)_{L}\otimes
SU(4|2)_{R}\otimes U(1)_{V}$ symmetry in the chiral limit and  this symmetry is
assumed to be spontaneously broken down to $SU(4|2)_{V}\otimes U(1)_{V}$
so that a connection with QCD can be made. The valence quark masses
and ghost quark masses are identical so that 
only the sea quarks contribute to the internal fermion loops (not coupled to 
electroweak gauge fields). 
If the ghost quark masses are set equal to infinity,
the theory becomes the two-flavor quenched theory, but more importantly,
if the sea quark masses
are reduced to those of the valence quark masses, then QCD is
recovered. 
In PQ$\chi $PT, nucleons are embedded in a {\bf 70} dimensional
representation ${\cal B}_{ijk}$ while the $\Delta$-resonances are
embedded in a {\bf 44} dimensional representation 
${\cal T}_{ijk}$, where the indices on both tensors run from $i,j,k=1$ to 
$6$. 
Therefore, at LO in PQ$\chi$PT
the isoscalar operators that contribute to the matrix elements of interest
have the form
\begin{eqnarray}
& & \overline{N}\ \Gamma\  N\rightarrow 
\overline{\cal B}^{ijk}\ \Gamma\  {\cal B}_{ijk}
\ \ , \ \ 
\overline{T}^{\alpha }\ \Gamma \ T_{\alpha }\rightarrow 
\overline{\cal T}^{\alpha,ijk}\ \Gamma\  {\cal T}_{\alpha, ijk}
\ \ \ , 
\end{eqnarray}
where $\Gamma =1$ or $S^{\mu }$.
While it is true that the extension of the flavor structure of the strange 
operator from QCD to PQQCD is not unique, we have extended the 
flavor singlet under SU(2) to a singlet under SU(4$|$2) for simplicity.
One could consider an extension to non-singlet operators, as discussed in
Ref.~\cite{PQmartin,GP01a}, 
but in this case such an extension has no obvious
advantages.

From a practical point of view, the strategy we are suggesting is that
simulations are performed with an unquenched $s$-quark, and two
partially-quenched non-degenerate light quarks.  This corresponds to 
PQQCD with SU(5$|$2) flavor symmetry.  
In order to perform the chiral extrapolation in the $u$ and $d$
quark masses one then matches the result of the SU(5$|$2) simulation onto
SU(4$|$2) PQ$\chi$PT.
However, in order to make use of the 
hierarchy between the $u$, $d$ and the $s$ quark masses, both the valence and
sea quark masses of the two 
light quarks must be much less than the mass of the 
$s$-quark. 
This approach should converge quite nicely for matrix elements in the nucleon 
for small enough meson masses, however, this is not expected to converge well
for matrix elements in baryons containing one or more $s$-quarks.
This is because the energy difference
between an intermediate state with a kaon, in three-flavor $\chi$PT,
e.g. $\Sigma\rightarrow N \overline{K} \rightarrow \Sigma$,
and an intermediate state with a pion,
e.g. $\Sigma\rightarrow \Sigma\pi \rightarrow \Sigma$,
is not large.

\subsection{Spin Singlet Operators}

The spin-singlet matrix elements we studied in the previous sections 
have the same chiral structures at NNLO. 
Since we have non-degenerate light quark masses, the
$s$-quark matrix elements will be different for protons and neutrons at
NNLO, but we will only consider the proton matrix elements here. 
We use the notation 
$\left\langle {\cal O}_{(2I+1,2S+1)}\right\rangle _{p}$ to denote proton matrix
elements of an operator with isospin $I$ and spin $S$.
Thus proton matrix elements of an isoscalar, spin-singlet operator,
such as $\left\langle x^{n-1}\right\rangle_{s,p}$ 
and $\langle p|\overline{s}s|p\rangle $ are denoted by 
$\left\langle {\cal O}_{(1,1)}\right\rangle _{p}$. 
In PQ$\chi $PT, up to NNLO, we find
\begin{eqnarray}
\left\langle {\cal O}_{(1,1)}\right\rangle_{p}^{PQ} & = &
\left\langle {\cal O}_{(1,1)}\right\rangle_{p}^{0}
\ -{\frac{g_{\Delta N}^{2}}{48 \pi ^{2}f^{2}}}
\left( \left\langle {\cal O}_{(1,1)}\right\rangle _{p}^{0}-\left\langle 
{\cal O}_{(1,1)}\right\rangle _{\Delta }^{0}\right) F_{0}  \nonumber \\
&&
+\ C_{S}^\prime\frac{\Delta }{\Lambda }
\ +\ D_{S}^\prime\frac{\Delta ^{2}}{\Lambda ^{2}}
\ +\ E_{S,1}^\prime\frac{m_{u}}{\Lambda}\ +\ E_{S,2}^\prime\frac{m_d}{\Lambda}
\ +\ E_{S,3}^\prime {m_j+m_l\over\Lambda}
\ \ ,  
\label{PQ1}
\end{eqnarray}
where $m_{j}$ and $m_{l}$ are sea quark masses and the function $F_0$ is
\begin{equation}
F_{0}=5J_{ud}+J_{uu}+J_{ju}+J_{lu}+2J_{jd}+2J_{ld}+2{\cal T}_{\eta _{u},\eta
_{u}}+2{\cal T}_{\eta _{d},\eta _{d}}-4{\cal T}_{\eta _{u},\eta _{d}}
\ \ \ .
\end{equation}
The functions $J_{ab}$ and ${\cal T}_{ab}$ 
are $J_{ab}=J(m_{ab},\Delta ,\mu )$ and $
{\cal T}_{ab}={\cal H}_{ab}(J_{a},J_{b},J_{X})$, with
\begin{eqnarray}
{\cal H}_{ab}(A,B,C) &=&-{\frac{1}{2}}\left[ \ {\frac{(m_{jj}^{2}-m_{\eta
_{a}}^{2})(m_{ll}^{2}-m_{\eta _{a}}^{2})}{(m_{\eta _{a}}^{2}-m_{\eta
_{b}}^{2})(m_{\eta _{a}}^{2}-m_{X}^{2})}}\ A-{\frac{(m_{jj}^{2}-m_{\eta
_{b}}^{2})(m_{ll}^{2}-m_{\eta _{b}}^{2})}{(m_{\eta _{a}}^{2}-m_{\eta
_{b}}^{2})(m_{\eta _{b}}^{2}-m_{X}^{2})}}\ B\right.  \nonumber \\
&&\left. \qquad \ +\ {\frac{(m_{X}^{2}-m_{jj}^{2})(m_{X}^{2}-m_{ll}^{2})}{
(m_{X}^{2}-m_{\eta _{a}}^{2})(m_{X}^{2}-m_{\eta _{b}}^{2})}}\ C\ \right] \ \
\ ,
\end{eqnarray}
and the tree-level meson 
masses are $m_{xy}^{2}\ =\lambda \left( m_{x}+m_{y}\right) $, 
$m_{\eta _{u}}^{2}\ =\ m_{uu}^{2}$ and $m_{X}^{2}=1/2(m_{jj}^{2}+m_{ll}^{2})$.
It is important to note that the low-energy constants $C_S$, $D_S$ and the 
$E_{S,i}$ are related to those of $\chi$PT, as can be seen by considering the 
limit $m_j\rightarrow m_u$ and $m_l\rightarrow m_d$.\, i.e. 
$C_S=C_S^\prime$, $D_S=D_S^\prime$, and 
$E_S=E_{S,1}^\prime+E_{S,2}^\prime + 2 E_{S,3}^\prime$.

\subsection{Spin Non-Singlet Operators}

The observables
$\mu _{s}$, $\left\langle
x^{n-1}\right\rangle _{\Delta s}$ and $\left\langle x^{n-1}\right\rangle
_{\delta s}$ are matrix elements of operators of the form
$\left\langle {\cal O}_{(1,3)}\right\rangle_p $. 
Their chiral expansions, up to NNLO in PQ$\chi $PT, are 
\begin{eqnarray}
\label{PQ2}
& & \left\langle {\cal O}_{(1,3)}\right\rangle _{p}
\ =\ \ \left\langle {\cal O}_{(1,3)}\right\rangle _{p}^{0}
\ +\ C_{V}\frac{\Delta }{\Lambda }
\ +\ D_{V}\frac{\Delta ^{2}}{\Lambda ^{2}}
\ +\ E_{V,1}\frac{m_u}{\Lambda}
\ +\ E_{V,2} \frac{m_d}{\Lambda}+E_{V,3}\frac{m_j+m_l}{\Lambda}
\\
& & \ -\ {\frac{1}{12\pi ^{2}f^{2}}}
\left[ \left(
g_{A}^{2}F_{1}
\ +\ g_{1}g_{A}F_{2}
\ +\ \frac{1}{4}{g_{1}^{2}F}_{3}
\ +\ \frac{1}{4} g_{\Delta N}^{2}F_{0}
\right) 
\left\langle {\cal O}_{(1,3)}\right\rangle_{p}^{0}
- \frac{5}{12} g_{\Delta N}^{2}F_{0}
\left\langle {\cal O}_{(1,3)}\right\rangle_{\Delta }^{0}
\right]
\ , 
\nonumber
\end{eqnarray}
where the loop functions are 
\begin{eqnarray*}
F_{1} &=&L_{ud}+L_{uu}+2L_{ju}+2L_{lu}+3R_{\eta _{u},\eta _{u}} \\
F_{2} &=&2L_{uu}-L_{ud}+L_{ju}+L_{lu}+3R_{\eta _{u},\eta _{u}}+3R_{\eta
_{u},\eta _{d}} \\
F_{3} &=&L_{uu}-5L_{ud}+2L_{lu}+3L_{ld}+2L_{ju}+3L_{jd}+3R_{\eta _{u},\eta
_{u}}+6R_{\eta _{u},\eta _{d}}+3R_{\eta _{d},\eta _{d}} 
\ \ \ ,
\end{eqnarray*}
with $L_{ab}=m_{ab}^{2}\log \left( {m_{ab}^{2}/\mu ^{2}}\right) $, 
and $R_{x,y}={\cal H}(L_{x},L_{y},L_{X})$.
The counterterms $C_V$, $D_V$, and the $E_{V,i}$ are related to 
the appropriate counterterms in $\chi$PT in a straightforward way.
Further, the $s$-quark contribution to the proton spin is
\begin{eqnarray}
\label{PQ3}
& & J_{s,p}\ =\ 
\ J_{s,p}^{0}
\ +\ C_{J_S}^\prime\frac{
\Delta }{\Lambda }
\ +\ D_{J_S}^\prime\frac{\Delta ^{2}}{\Lambda ^{2}}
+E_{J_S,1}^{\prime }\frac{m_u}{\Lambda}+E_{J_S,2}^{\prime }\frac{
m_d}{\Lambda}+E_{J_S,3}^{\prime }\frac{m_j+m_l}{\Lambda}
\\ 
&&\ -\ {\frac{1}{12\pi ^{2}f^{2}}}\left[ (J_{s,p}^{0}-
\frac{\left\langle x\right\rangle _{s,\pi }^{0}}{2})\left(
g_{A}^{2}F_{1}+g_{1}g_{A}F_{2}+\frac{1}{4}{g_{1}^{2}F}_{3}+\frac{1}{4}
g_{\Delta N}^{2}F_{0}\right) 
\right.
\nonumber\\
& & \left.
\qquad\qquad -\frac{5}{12}(J_{s,\Delta }^{0}-\frac{\left\langle
x\right\rangle _{s,\pi }^{0}}{2})g_{\Delta N}^{2}F_{0}
\right]
\ ,
\nonumber 
\end{eqnarray}
where the counterterms in  eq.~(\ref{PQ3}) are related to the counterterms in
$\chi$PT, given in eq.~(\ref{eq:qcdspin}), by
$C_{J_s}= C_{J_s}^\prime$, $D_{J_s}= D_{J_s}^\prime$, and 
$E_{J_S} = E_{J_S,1}^{(\prime)}+E_{J_S,2}^{(\prime )}+2 E_{J_S,3}^{(\prime )}$.

\section{Conclusions}

The chiral extrapolation of matrix elements computed in lattice QCD from the
lattice quark masses to their physical values will play a central role in
future lattice simulations.
The slow convergence properties of
three-flavor $\chi$PT will not impact the chiral extrapolation of strange
matrix elements in the nucleon when simulations can be performed with
pions of mass $\lesssim 300~{\rm MeV}$.
Two-flavor $\chi$PT is sufficient for the chiral extrapolation of
these matrix elements. 
We have presented the chiral expansion of several
strange matrix elements at NNLO in both $\chi$PT and PQ$\chi$PT.


\acknowledgements
We would like to thank Silas Beane for useful discussions.
MJS would like to thank the Benasque Center for Science in Benasque, 
Spain for providing a nice environment and for kind hospitality
during some of this work. 
JWC is supported in part by the U.S. Dept. of Energy under grant No.
DE-FG02-93ER-40762. MJS is supported in part by the U.S. Dept. of Energy
under grant No. DE-FG03-97ER-41014.

\end{document}